# Numerical modelling of *Ar* glow discharge at intermediate and high pressures

G. M. Petrov and C. M. Ferreira

Centro de Electrodinâmica da Universidade Técnica de Lisboa,

Instituto Superior Técnico, 1096 Lisboa Codex, Portugal

**Abstract**

A detailed collisional-radiative model for *Ar* at intermediate and high pressures is developed. The model is coupled with the electron Boltzmann equation and includes several highly excited *Ar* states and charged particles. The densities of all neutral and charged species considered are calculated self-consistently. Detailed information concerning the electronic data used is presented. The model will be further used for analysing the contraction phenomena in noble gas discharges and eventually for theoretical description of the kinetic properties of high pressure *Ar* excimer laser.

# 1. Introduction.

In recent years the interest in high pressure gas discharges substantially increases. They can be used in plasma processing, light sources, elemental analysis etc. Collisional-radiative models of gas discharges constitute a powerful theoretical methods for self-consistent calculations of a number of physical quantities, necessary for a complete understanding of the plasma properties. Our intention is to developed a detailed collisional-radiative model for *Ar* plasma at intermediate and high pressures and to provide electronic data for it. The model can be used for modelling high pressure *Ar* discharges and can provide an important insight of several discharge phenomena, such as contraction of a DC positive column and Surface Wave Discharges. The transition from glow discharge to arc and the deviation from LTE can be analysed too. An important application of the model could be the theoretical description of the kinetic properties of high pressure *Ar* excimer laser, operating in the UV region. Such a laser is undoubtedly of great interest due to its unique properties and the model developed is an important step in predicting and modeling the discharge conditions at which this laser operates.

At pressures above about 10 Torr the discharge kinetics is much more complicated compared to low pressure discharges. The difficulties arise mainly when trying to take into account the three-body collision processes. With the gas pressure increasing their variety and importance increases too. While the conversion of atomic to molecular ions is important even at low pressures, at high pressures new charged and neutral species become playing important role. These are for example $Ar_3^+$ and $Ar_2^*$. An increase of the gas pressure causes nonlinear increase of both the concentration of $Ar_3^+$ and the population of $Ar_2^*$. At certain experimental conditions the population of $Ar_2^*$ could reach values of $10^{15}$ $cm^{-3}$, which is enough to create powerful laser generation in the UV region. Due to the low mean energy and high electron density the three-body recombination processes gain importance changing significantly the discharge properties. Another feature distinguishes the high pressure from low pressure discharges. While at low pressures a combination of parameters can be made, such as *pR*, *E/N*, $n_e/N$ and $N_k/N$, where $N_k$ are the populations of the excites states or charged particles, *p* is the pressure, *E* is the electric field and *R* is the tube radius, at high pressures such combinations are not possible due to the nonlinear character of the processes and the discharge properties depend on each parameter separately.

At high pressures the populations of the excited states increase and the trapping of radiation for allowed radiative transitions becomes considerable. The effective lifetime of the highly excited states decrease and the impact of the electron impact excitation and deexcitation increase which tend to bring these states into equilibrium. Their importance as electron production source increase and they must be taken into account too.

Various collisional-radiative models for *Ar* plasma have been developed in the past [1-16]. Some of them are applicable at different discharge conditions, usually at low pressures [1-7]. There are models for high pressure *Ar* discharges too, which are, however, designed for particular problems and can not describe properly all plasma parameters [8-11]. Simple analytical models with one lumped excited state only has also been developed [12-14]. This article deals with a detailed analysis of the electron kinetics in pure argon at intermediate and high pressures based on solution of the homogeneous electron Boltzmann equation and the balance equations for all species considered. A complete set of cross-sections and rate constants is compiled, which is presented in series of tables and figures. The model allows a detailed analysis of the electron kinetic properties (Electron Energy Distribution Function (EEDF), transport parameters, rate coefficients for excitation and ionisation, fractional particle and energy gain and losses) to be made.

**2. Kinetic model.**

An accurate description of the discharge properties is possible if only an appropriate set of neutral and charged species coupled with the electron Boltzmann equation is considered. The following species are involved in the kinetic model: *Ar*, *Ar(4s)*, *Ar(4p)*, *Ar(3d)*, *Ar(5s)*, *Ar(5p)*, *Ar(4d)*, *Ar(6s)*, $Ar_2^*$, $Ar^+$, $Ar_2^+$ and $Ar_3^+$. The energy level diagrams for *Ar* and $Ar_2^*$ are taken from [1, 11] (Fig.1). All excited states are considered as blocks of levels. It is justified since both the atom and the electron densities are high and the transitions between the levels in each group tend to bring them into equilibrium. The molecular dimmer $Ar_2^*$ has been taken into account, since it is formed by three-body collisions of *Ar(4s)* with two *Ar* atoms and at high pressures its density could be considerable. The effective energy of each block is taken as the energy of the center, i.e. $E_k = \sum_{l=1}^{m} E_{k,l}/m$, where $E_{k,l}$ is the energy of each level in the block $k$, accounted from the ground state and $m$ is their number. In all calculations the averaged populations and *g*-factors of the excited states are used, i.e. $n_k = \sum_{l=1}^{m} n_{k,l}/m$ and $g_k = \sum_{l=1}^{m} g_{k,l}/m$, where $n_{k,l}$ and $g_{k,l}$ are the population and *g*-factor of each level in the block respectively.

Eighty seven plasmo-chemical reactions are taken into account in the numerical code, whose rate constants are either taken from other sources, or calculated in this work. These include: elastic scattering, diffusion, excitation and deexcitation processes between the ground and excited *Ar* states, all allowed radiative transitions between the excited *Ar* states, chemi-ionization, three-body and dissociative recombination, conversion to molecular ions and other processes. All reactions are listed in Tables 1-5. The gas temperature dependence is also taken into account in dissociative recombination and heavy-particle collision processes. In addition, quasineutrality is assumed, which in our case reads $n_e = Ar^+ + Ar_2^+ + Ar_3^+$.

It is well recognised, that in inert gases the two-term expansion gives sufficient accuracy in obtaining the discharge properties and the electron Boltzmann equation has been solved using two-term expansion in Legendre polynomials. The EEDF is normalized by $\int_0^\infty u^{1/2} f^0(u) du = 1$.

Several ionization processes are considered in the model: ionization from both the ground and the excited states, associative ionization and Penning ionization. All these processes are treated as processes in which electrons are created. It is assumed that after an ionization event the primary and the ejected electrons share the remained energy. It has been shown that such an approach gives sufficient accuracy in calculating the EEDF [17]. The loss of electrons is through recombination and diffusion processes. Both dissociative and three-body recombinations are considered. The diffusion loss term is described by generalized diffusion frequency $\nu_{diff}$. More details will be given in the next section.

It is well known that in inert gases the electron-electron collisions can strongly affect the EEDF even at degree of ionization about $10^{-7}$ - $10^{-6}$ [3, 6]. We took into account in the electron Boltzmann equation not only the electron-electron collisions, but the electron-ion collisions too. The latter can not be neglected since at high pressures the mean energy is very low and the energy loss in electron-neutral collisions rapidly decreases due to the Ramsauer minimum, while the energy loss in electron-ion collisions increases with the mean energy decreasing.

The populations of the excited states strongly depend on the radiation processes. The trapping of radiation is substantial not only for the resonance levels, but for all excited states too. This is especially prominent at high pressures, since the excited states are well populated and close to equilibrium. The trapping of radiation has been taken into account in all allowed radiative transitions considered.

The applicabilty of the model should be further discussed. Since the model do not consider each level in the *4s* block of levels separately, the accuracy at low pressures (below 10 Torr) is limited. The model provides an accurate description of the discharge properties for pressures above about 10 Torr.

**3. Atomic data.**

In this section details about all atomic data used in our model are given. Some of the cross-sections are calculated, other are taken from different sources.

*3.1 Elastic scattering*

The elastic scattering of electrons with inert gases is well studied both theoretically and experimentally. Several techniques has been applied to measure these cross-sections with increasing precision and there are a lot of data for *Ar* in the literature [18-20]. A summary is given in a recent article [20]. The cross-section used in our model is taken from [18].

*3.2 Electron impact excitation and deexcitation*

Though *Ar* has been extensively studied, only few electron impact excitation cross-sections are known experimentally. The studies are limited to excitation from the ground state [18, 21-25] and few transitions from *4s* to *4p* blocks of levels. We have used experimentally measured cross-sections for excitation of the *4s* and *4p* blocks of levels from the ground atom state [18]. Compilation of these cross-sections is given in [6, 15]. With respect to the *5p* block of levels a semi-empirical formula is used [15]. There are no available cross-sections for transition between the excited states and for optically allowed transitions (s ⇔ p and p ⇔ d )the cross-sections are computed using the well known Drawin's formula [26] (excitation cross-sections for optically forbidden transitions are not considered). This formula is also used for computing the excitation cross-sections from the ground atom state to the *5s, 6s, 3d* and *4d* blocks of levels. Semi-empirical approximations for excitation from the ground state is given in [15]. The oscillator strength used is a sum of the oscillator strengths of all allowed transitions between each state of the lower block of levels (or ground state) and the upper block of levels. Thus we present the total cross-section for electron impact excitation, i.e. the sum of all cross-sections between the corresponding blocks of levels.

The excitation cross-section used in our caculations is in the form $\sigma_{exc}(u)=\sigma_0(\Delta E)\chi(u/\Delta E)$, where $\chi(x) = \frac{x-1}{x^2}\ln(1.25x)$, $u$ is the incident electron energy and $E_{mn}$ is the energy difference between the two blocks of levels. Both $\Delta E$ and the parameter $\sigma_0$ are given in Table 1. In addition, the oscilator strengths and the *g*-factors $g_l$ and $g_u$ for the lower block of levels (or ground state) and for the higher block of levels respectively are given. The *g*-factors are necessary for calculation of the corresponding deexcitation cross-sections. The oscillator strengths for transition from the ground state to the *5s*, *6s*, *3d* and *4d* blocks of levels have been taken from [27] and for all other transitions - from [28]. A recent compilation is made in [29]. The energy difference is taken as the energy between the centers of both blocks except the transitions from the ground state to *4s* and *4p* levels in which the energy difference is accounted from the ground state and the lowest lying level in the group. Some of the cross-sections are computed and plotted in [30, 31] and analytical formula is given in [9]. The cross-sections for excitation from the ground state are tabulated in Table 6 and plotted in Fig. 2; all others - in Table 7 and Fig. 3 respectively.

The cross-section for excitation from *4s* to *4p* block of levels is of utmost interest. There are several attempts to investigate it, however, the cross-sections for few single transitions only are known [30, 31]. An averaged cross-section is plotted in [30, 31]. We have used the cross-section calculated in this work.

All cross-sections for electron impact deexcitation $\sigma_{deexc}$ are deduced from the corresponding excitation cross-section using the method of detailed balancing [9], which reads: $\sigma_{deexc}(u)=\sigma_{exc}(u+\Delta E)(u+\Delta E)g_l/(ug_u)$,

*3.3 Electron impact ionization*

The cross-section for ionization from the ground atom state has been taken from [18]. This cross-section in a wide energy interval is also tabulated in [32]. The ionization cross-sections from all other *Ar* excited states as well as from $Ar_2^*$ has been calculated using a simple analytical formula given by Vriens [33]. An alternative formula is given in [26]. Vriens' formula is applicable for excited atom states and fits well the available experimental data for noble gases, including *Ar* [33]. The cross-sections for ionisation from *4s* and *4p* blocks of levels have been calculated in [34, 35] and also given in [15]. Several other are plotted in [9].

The ionisation cross-section can be written in the form $\sigma_{ion}(u)=\sigma_0(E_{pi})\varphi(u/E_{pi})n_p$, where $\varphi(x) = \dfrac{5/3 - 1/x - 3/x^2}{3.25 + x}$, $E_{pi}$ is the ionization energy accounted from the center of the corresponding block of levels and $n_p$ is the number of levels in the block. Both $E_{pi}$ and the parameter $\sigma_0$ are given in Table 2. Similar to the excitation cross-sections, the total cross-sections for ionization are calculated and further used. They are tabulated in Table 8 and plotted in Fig. 4.

*3.4 Recombination*

With increasing pressure the importance of the diffusion losses rapidly decreases and the main loss particle channel is the recombination. The recombination processes are extensively studied both theoretically [36] and experimentally [37, 38]. We have considered eight recombination processes, listed in Table 3.

The dissociative recombination has been studied experimentally by several authors [39-46] and the rate constants given in different sources are in good agreement. The electron temperature dependence is taken into account in most of them, but few authors only have investigated the impact of the gas temperature. At high pressure the gas temperature may deviate from the room temperature and the gas temperature dependence must be taken into account too. In our calculations we used rate constant taken from [41].

It is generally assumed that the dissociative recombination produces *Ar(4p)* and an *Ar* atom in a ground state. Spectroscopic measurements justify such an assumption [46]. Other measurements show, however, that the recombination flux populates predominantly the *4s* states, not the *4p* states [45]. We have assumed that after a recombination event the excited *Ar* atom is in *4p* state.

The most problematic reaction is $Ar_k^+ + Ar + e \Rightarrow products$ with $k = 1,2,3$, whose rate constant depend on both the electron and gas temperatures. It has been estimated for example in [47]. We have calculated it using an empirical formula provided in [36].

*3.5 Heavy-particle collisions*

All heavy-particle collision processes found in the literature are considered. It includes chemi-ionisation processes, conversion to molecular ions and $Ar_2^*$ and transitions between the excited $Ar$ states. Their rate constants have been taken from several sources [48-55]. Most of them are summarized in [10, 11]. Both the processes and the corresponding rate constants are listed in Table 4. The gas temperature dependence, where available, is taken into account.

One of the most important reactions is the conversion of atomic to molecular ion $Ar^+ + Ar + Ar \Rightarrow Ar_2^+ + Ar$ since it creates molecular ion which consequently recombine. The rate constant has been measured by several authors and with the exception of that in [40] they are in good agreement. We have adopted the rate constant measured in [51].

*3.6 Radiative transitions*

All allowed radiative transitions are considered in the model (Table 5). In each process the trapping of radiation is taken into account both for the resonance and excited states [56, 57]. Since we use blocks of levels, the transition probability is taken as a sum of the transition probabilities of all allowed transitions. The electronic data are taken from [27-29]. Since for several transitions electronic data are missing, the oscillator strengths and transition probabilities are calculated, using Kramers formula [36].

*3.7 Diffusion*

The diffusion of all ion species is considered in the model. The ambipolar diffusion frequency is $v_{diff} = \left(\frac{2.4}{R}\right)^2 D_a$, where $D_a = \dfrac{b_e \sum_{k=1}^{3} \alpha_k D_{i,k} + D_e \sum_{k=1}^{3} \alpha_k b_{i,k}}{b_e + \sum_{k=1}^{3} \alpha_k b_{i,k}}$ is the ambipolar diffusion coefficient.. Here $b_e$ is the electron mobility and $b_{i,k}$ and $D_{i,k}$ are the ion mobility and diffusion coefficient respectively for the $k$-th type of ion and $\alpha_k$ is the fraction of each ion. The electron mobility has been determined using the calculated EEDF. The mobilities of $Ar^+$ and $Ar_2^+$ have been taken from [58]. They are $b_{Ar^+} = 1.535 \ cm^2V^{-1}s^{-1}$ and $b_{Ar_2^+} = 1.833 \ cm^2V^{-1}s^{-1}$ at $p = 1$ atm. and $T_g$=300 K respectively. We did not find any data for $Ar_3^+$ mobility and we assumed that it is equal to the mobility of $Ar_2^+$.

**4. Results and discussion.**

Figures 5-8 illustrate the discharge properties of high pressure *Ar* DC glow discharge at 100 Torr pressure and 1 cm inner tube diameter. The discharge current is up to 6 mA. At higher currents the discharge contracts, the radial distribution of the electron density is no longer Bessel function (or close to Bessel function) and strict calculations of its radial properties must be made. In order to illustrate the validity of the model, a comparison with experimental results taken from [59] is made. The gas temperature, which is an input parameter in our code, has been taken from the same experiment.

Fig. 5 displays the EEDF, calculated at discharge currents $i = 1$ mA and $i = 6$ mA. Since the degree of ionization is very small, the electron-electron collisions has minor importance and the EEDF changes only slightly in this discharge current range.

The dependence of the axial electric field and the electron density and electron temperature at the discharge axis of the discharge current is shown in this figure. With the discharge current increasing the electron density at the discharge axis increases proportionally. The calculated electron density at the discharge axis is in excellent agreement with the experimentally measured values (Fig.6a). The discrepancy with respect to the electron temperature is higher; about 20 per cent (Fig.6b). Probably the electron temperature has been measured assuming Maxwellian EEDF, while the EEDF is close to Druwestain (Fig.5). This could be an explanation why the calculated electron temperature is lower than the experimentally measured one. Both the calculated and the experimentally measured electric field show similar tendency; they decrease with the discharge current increasing. The discrepancies between our calculations and the experiment are about 10 per cent (Fig.6c).

Figs. 7 and 8 show several other discharge characteristics: the ion densities (Fig.7a), the populations of *Ar(4s)* and *Ar(4p)* blocks of levels and $Ar_2^*$ state (Fig.7b), electron mobility and diffusion coefficient (Fig.8a) and the absorbed energy per electron $\theta / N$ (Fig.8b). The population of *Ar(4p)* block of levels is about three orders of magnitude smaller than the population of *Ar(4s)* block of levels. The populations of the other blocks of levels is negligible. However, if the discharge current increases, the electron density increases too and the population of *Ar(4p)* and all other higher lying blocks of levels also increases. Due to the stepwise character of the transitions between these blocks of levels such an increase is strongly nonlinear and at electron densities at the discharge axis $10^{12}$ *cm*$^{-3}$ - $10^{13}$ *cm*$^{-3}$ these blocks of levels would be in equilibrium with respect to the electron temperature.

The results presented so far refer to very low discharge currents. At higher currents the discharge contracts and its properties differ significantly from discharge in a diffuse state. With the model presented we have investigated a contracted DC glow discharge and the transition from diffuse to a contracted state. This requires to solve a coupled system of fluid equations together with the electron Boltzmann equation. The numerical technique is given in details in [60].

The next figure presents the main plasma parameters (the electron density, electron and gas temperature at the discharge axis and the electric field) as a function of the discharge current at the same conditions as Figs. 5-8. For $i < 6$ mA the discharge is in a diffuse state and with the current increasing the discharge abruptly goes to a contracted state. The transition can be clearly seen in Fig. 9. The results obtained with our model have been compared with experiment. Unilke the diffusion discharge with the discharge current increasing the electron density at the discharge axis increase nonmonotoneously due to the contraction. The gas temperature increase, while both the electric field and the electron temperature, which is defined as two-third of the mean energy, decrease. The overall comparison show that the theoretical and experimental results are in good agreement. Another comparison with experiment is presented in Fig.10, but for tube radius $R$=1.3 cm, gas pressure $p$=20 Torr and different discharge current range [61]. As in Fig. 9, the theoretical results are close to the experimental.

All results in Figs. 9 and 10 have been obtained accounting for the radial structure of the EEDF and all other quantities [60]. It includes calculation of the radial distribution of the electron and atom densities and the gas temperature and the solution is as a whole very sensitive and can accumulate errors. The experimental errors in [57, 59] have to be taken into account too.

## 5. Conclusion

A steady-state collisional-radiative model for *Ar* in cylindrical geometry at intermediate and high pressures is developed. The model is based on simultaneous solution of the electron Boltzmann equation and rate balance equations for several excited *Ar* states and charged particles. Both the EEDF and the densities of all species considered are calculated self-consistently with high precision. The model considers *Ar(4s)*, *Ar(4p)*, *Ar(3d)*, *Ar(5s)*, *Ar(5p)*, *Ar(4d)*, *Ar(6s)* excited atomic states, one excited molecular state ($Ar_2^*$) and three ion species ($Ar^+$, $Ar_2^+$ and $Ar_3^+$).

We have carried out an extensive search of electronic data for elementary processes and detailed information concerning the electronic data included in the model is presented. A complete set of cross sections for electron impact excitation and ionisation is tabulated. Rate constants for recombination and heavy-particle collisions are compiled from different sources and compared.

Calculations are carried out and compared with experiment. The electron density, electron temperature and the electric field are very sensitive quantities and we have tested our model based on these quantities. The experimental values and that obtained with the model are in good agreement both for diffuse and contracted discharge.

The model could be further improved if all four lowest lying levels are considered separately. It would make the model more flexible and applicable at low pressures.

**Figure captions**

Fig.1. Energy level diagrams of *Ar* and $Ar_2$.

Fig.2. Absolute partial electron-impact excitation cross sections of *Ar* from the ground state as a function of the kinetic energy.

Fig.3. Absolute partial electron-impact excitation cross sections of *Ar* between the excited states as a function of the kinetic energy.

Fig.4. Absolute partial electron-impact ionisation cross sections of *Ar* from the ground state and from the excited atomic and molecular states as a function of the kinetic energy.

Fig.5. Calculated EEDF at gas pressure $p$ = 100 Torr, tube radius $R$ = 1cm and gas temperature $T_g$ = 300 K for discharge currents $i$ = 1 mA and $i$ = 6 mA.

Fig.6. Dependence of the electron density (a), electron temperatute (b) and the electric field (c) on the discharge current at the conditions in Fig.5. Full cycles are experimental results from [59].

Fig.7. Dependence of the ion densities (a) and the populations of *Ar(4s)*, *Ar(4p)* blocks of levels and $Ar_2^*$ level (b) on the discharge current at the conditions in Fig.5.

Fig.8. Dependence of the electrom mobility $b_e$ and diffusion coefficient $D_e$ (a) and the mean energy absorbed per electron (b) on the discharge current. The conditions are the same as in Fig.5.

Fig.9 Dependence of the electron density (a) and electron and gas temperatures (b) at the discharge axis and the electric field (c) on the discharge current. The gas pressure is 100 Torr and the tube radius is 1 cm.

Fig.10 The same as in Fig.9 for gas pressure 20 Torr and the tube radius is 1.3 cm.

**Table captions**

Table 1. Reactions for electron impact excitations considered in the model and parameters for calculations of the excitation and deexcitation cross sections.

Table 1. Reactions for electron impact ionisations considered in the model and parameters for calculations of the cross sections.

Table 3. Reactions and rate coefficients for recombinations considered in the kinetic model.

Table 4. Reactions and rate coefficients for heavy-particle collisions considered in the kinetic model.

Table 5. Spontaneous emission reactions and transition probabilities.

Table 6. Cross sections for excitation from the ground state by electron-impact.

Table 7. Cross sections for transitions between excited state by electron-impact.

Table 8. Cross sections for ionisation by electron-impact

Table 1

| process | ΔE (eV) | f | $g_l$ | $g_u$ | $\sigma_0$ (cm$^2$) | reference |
|---|---|---|---|---|---|---|
| Ar + e ⇒ Ar (4s) + e | 11.55 | ___ | 1 | 3 | ___ | [18] |
| Ar + e ⇒ Ar (4p) + e | 12.91 | ___ | 1 | 3.6 | ___ | [18] |
| Ar + e ⇒ Ar (3d) + e | 14.1 | $2.03 \times 10^{-1}$ | 1 | 5 | $6.65 \times 10^{-17}$ | calculated |
| Ar + e ⇒ Ar (5s) + e | 14.1 | $4.04 \times 10^{-2}$ | 1 | 3 | $1.32 \times 10^{-17}$ | calculated |
| Ar + e ⇒ Ar (5p) + e | 14.6 | ___ | 1 | 3.6 | ___ | [15] |
| Ar + e ⇒ Ar (4d) + e | 14.8 | $6.70 \times 10^{-2}$ | 1 | 5 | $1.99 \times 10^{-17}$ | calculated |
| Ar + e ⇒ Ar (6s) + e | 14.8 | $3.14 \times 10^{-2}$ | 1 | 3 | $9.33 \times 10^{-18}$ | calculated |
| Ar(4s) + e ⇒ Ar (4p) + e | 1.51 | 4.24 | 3 | 3.6 | $1.21 \times 10^{-13}$ | calculated |
|  |  | ___ | ___ |  | $5.52 \times 10^{-14}$ | [9] |
| Ar(4s) + e ⇒ Ar (5p) + e | 2.91 | $4.06 \times 10^{-2}$ | 3 | 3.6 | $3.12 \times 10^{-16}$ | calculated |
| Ar(4p) + e ⇒ Ar (3d) + e | 0.9 | 5.07 | 3.6 | 5 | $4.08 \times 10^{-13}$ | calculated |
| Ar(4p) + e ⇒ Ar (5s) + e | 0.9 | 1.84 | 3.6 | 3 | $1.48 \times 10^{-13}$ | calculated |
| Ar(4p) + e ⇒ Ar (4d) + e | 1.6 | $2.48 \times 10^{-1}$ | 3.6 | 5 | $6.31 \times 10^{-15}$ | calculated |
| Ar(4p) + e ⇒ Ar (6s) + e | 1.6 | $1.46 \times 10^{-1}$ | 3.6 | 3 | $3.71 \times 10^{-15}$ | calculated |
| Ar(3d) + e ⇒ Ar (5p) + e | 0.5 | $8.90 \times 10^{-2}$ | 5 | 3.6 | $2.32 \times 10^{-14}$ | calculated |
| Ar(5s) + e ⇒ Ar (5p) + e | 0.5 | 5.10 | 3 | 3.6 | $1.35 \times 10^{-12}$ | calculated |
| Ar(5p) + e ⇒ Ar (4d) + e | 0.2 | 25.4 | 3.6 | 5 | $4.13 \times 10^{-11}$ | calculated |
| Ar(5p) + e ⇒ Ar (6s) + e | 0.2 | 25.4 | 3.6 | 3 | $4.13 \times 10^{-11}$ | calculated |

Table 2

| process | $E_{pi}$ (eV) | $\sigma_0$ (cm$^2$) | reference |
|---|---|---|---|
| $Ar + e \Rightarrow Ar^+ + 2e$ | 15.76 | — | [18] |
| $Ar(4s) + e \Rightarrow Ar^+ + 2e$ | 4.07 | $1.57 \times 10^{-14}$ | calculated |
| $Ar(4p) + e \Rightarrow Ar^+ + 2e$ | 2.52 | $1.03 \times 10^{-13}$ | calculated |
| $Ar(3d) + e \Rightarrow Ar^+ + 2e$ | 1.66 | $2.84 \times 10^{-13}$ | calculated |
| $Ar(5s) + e \Rightarrow Ar^+ + 2e$ | 1.66 | $9.45 \times 10^{-14}$ | calculated |
| $Ar(5p) + e \Rightarrow Ar^+ + 2e$ | 1.16 | $4.84 \times 10^{-13}$ | calculated |
| $Ar(4d) + e \Rightarrow Ar^+ + 2e$ | 0.96 | $8.48 \times 10^{-13}$ | calculated |
| $Ar(6s) + e \Rightarrow Ar^+ + 2e$ | 0.96 | $2.83 \times 10^{-13}$ | calculated |
| $Ar_2^* + e \Rightarrow Ar_2^+ + 2e$ | 3.80 | $5.31 \times 10^{-15}$ | calculated |

| reaction | rate | ref. |
|---|---|---|
| $Ar_2^+ + e \Rightarrow Ar(4p) + Ar$ | $9.1 \times 10^{-7} \left( \dfrac{T_e(K)}{300} \right)^{-0.61} \dfrac{cm^3}{s}, T_g = 300 K$ | [39] |
| | $6.0 \times 10^{-7} \, cm^3 s^{-1}, \quad T_g = 300 K$ | [40] |
| | $8.5 \times 10^{-7} \left( \dfrac{T_e(K)}{300} \right)^{-0.67} \left( \dfrac{T_g(K)}{300} \right)^{-0.58} \dfrac{cm^3}{s}$ | [41] |
| | $8.0 \times 10^{-7} \left( \dfrac{T_e(K)}{300} \right)^{-0.67} \dfrac{cm^3}{s}, T_g = 300 K$ | [42] |
| $Ar_3^+ + e \Rightarrow Ar(4p) + 2Ar$ | $1.6 \times 10^{-7} T_e(eV)^{-0.54} cm^3 s^{-1}$ | [10] |
| $Ar^+ + e + e \Rightarrow Ar(4p) + e$ | $5.4 \times 10^{-27} T_e(eV)^{-4.5} cm^6 s^{-1}$ | [36] |
| $Ar_2^+ + e + e \Rightarrow Ar(4p) + Ar + e$ | $5.4 \times 10^{-27} T_e(eV)^{-4.5} cm^6 s^{-1}$ | [36] |
| $Ar_3^+ + e + e \Rightarrow Ar_2^* + Ar + e$ | $5.4 \times 10^{-27} T_e(eV)^{-4.5} cm^6 s^{-1}$ | [36] |
| $Ar^+ + Ar + e \Rightarrow Ar(4p) + Ar$ | $3.7 \times 10^{-29} T_e(eV)^{-1.5} T_g(K)^{-1} cm^6 s^{-1}$ | [36] |
| $Ar_2^+ + Ar + e \Rightarrow Ar(4p) + 2Ar$ | $3.7 \times 10^{-29} T_e(eV)^{-1.5} T_g(K)^{-1} cm^6 s^{-1}$ | [36] |
| $Ar_3^+ + Ar + e \Rightarrow Ar_2^* + 2Ar$ | $3.7 \times 10^{-29} T_e(eV)^{-1.5} T_g(K)^{-1} cm^6 s^{-1}$ | [36] |

Table 4

| reaction | rate | ref. |
|---|---|---|
| $Ar(4s) + Ar(4s) \Rightarrow Ar^+ + Ar + e$ | $1.2 \times 10^{-9} \left( \dfrac{T_g(K)}{300} \right)^{1/6} \dfrac{cm^3}{s}$ | [48] |
| | $5.0 \times 10^{-10} \, cm^3 s^{-1}$ | [10, 49] |
| $Ar(4p) + Ar(4p) \Rightarrow Ar^+ + Ar + e$ | $5.0 \times 10^{-10} \, cm^3 s^{-1}$ | [10, 49] |
| $Ar(4p) + Ar(4s) \Rightarrow Ar^+ + Ar + e$ | $5.0 \times 10^{-10} \, cm^3 s^{-1}$ | [10, 49] |
| $Ar(4s) + Ar(4s) \Rightarrow Ar_2^+ + e$ | $5.0 \times 10^{-10} \, cm^3 s^{-1}$ | [10, 49] |
| | $6.0 \times 10^{-10} \, cm^3 s^{-1}$ | [11] |
| $Ar^+ + Ar + Ar \Rightarrow Ar_2^+ + Ar$ | $9.4 \times 10^{-33} \left( \dfrac{T_g(K)}{300} \right)^{-0.27} \dfrac{cm^6}{s}, \quad 300 < T_g(K) < 1500$ | [40] |
| | $2.50 \times 10^{-31} \, cm^6 s^{-1}, \quad T_g = 300 \, K$ | [50] |
| | $2.25 \times 10^{-31} \left( \dfrac{T_g(K)}{300} \right)^{-0.4} \dfrac{cm^6}{s}, \quad 150 < T_g(K) < 300$ | [51] |
| | $2.07 \times 10^{-31} \, cm^6 s^{-1}, \quad T_g = 300 \, K$ | [52] |
| | $2.30 \times 10^{-31} \left( \dfrac{T_g(K)}{300} \right)^{-0.61} \dfrac{cm^6}{s}, \quad 77 < T_g(K) < 296$ | [53] |
| $Ar_2^+ + Ar + Ar \Rightarrow Ar_3^+ + Ar$ | $6.96 \times 10^{-32} \left( \dfrac{T_g(K)}{298} \right)^{-0.47} \dfrac{cm^6}{s}, \quad 77 < T_g(K) < 298$ | [54] |
| | $3.5 \times 10^{-30} \, cm^6 s^{-1}, \quad T_g = 77 \, K$ | [53] |
| $Ar_3^+ + Ar \Rightarrow Ar_2^+ + Ar$ | $8.65 \times 10^{-12} \left( \dfrac{T_g(K)}{298} \right)^{-0.73} \dfrac{cm^3}{s}, \quad 77 < T_g(K) < 298$ | [54] |
| | $9 \times 10^{-12} \, cm^3 s^{-1}$ | [10] |
| $Ar(4s) + 2Ar \Rightarrow Ar_2^* + Ar$ | $1.0 \times 10^{-32} \, cm^6 s^{-1}$ | [10, 11] |
| | $(1.0 \pm 0.3) \times 10^{-32} \, cm^6 s^{-1}$ | [55] |
| $Ar(4p) + 2Ar \Rightarrow Ar(4s) + 2Ar$ | $5.0 \times 10^{-32} \, cm^6 s^{-1}$ | [10] |
| $Ar(4p) + Ar \Rightarrow Ar(4s) + Ar$ | $5.0 \times 10^{-11} \, cm^3 s^{-1}$ | [10] |

Table 5

| reaction | A [1/s] | ref. |
|---|---|---|
| $Ar(4s) \Rightarrow Ar+h\nu$ | $7.00 \times 10^8$ | [27] |
| $Ar(5s) \Rightarrow Ar+h\nu$ | $1.17 \times 10^8$ | [27] |
| $Ar(6s) \Rightarrow Ar+h\nu$ | $7.75 \times 10^8$ | [27] |
| $Ar(3d) \Rightarrow Ar+h\nu$ | $5.96 \times 10^8$ | [27] |
| $Ar(4d) \Rightarrow Ar+h\nu$ | $2.26 \times 10^8$ | [27] |
| $Ar(4p) \Rightarrow Ar(4s)+h\nu$ | $3.76 \times 10^8$ | [28, 29] |
| $Ar(5p) \Rightarrow Ar(4s)+h\nu$ | $1.65 \times 10^7$ | [28, 29] |
| $Ar(3d) \Rightarrow Ar(4p)+h\nu$ | $1.46 \times 10^8$ | [28, 29] |
| $Ar(5s) \Rightarrow Ar(4p)+h\nu$ | $8.81 \times 10^7$ | [28, 29] |
| $Ar(4d) \Rightarrow Ar(4p)+h\nu$ | $2.70 \times 10^7$ | [28, 29] |
| $Ar(6s) \Rightarrow Ar(4p)+h\nu$ | $1.78 \times 10^7$ | [28, 29] |
| $Ar(5p) \Rightarrow Ar(3d)+h\nu$ | $1.38 \times 10^6$ | [28, 29] |
| $Ar(5p) \Rightarrow Ar(5s)+h\nu$ | $3.15 \times 10^7$ | calc. |
| $Ar(4d) \Rightarrow Ar(5p)+h\nu$ | $2.80 \times 10^7$ | calc. |
| $Ar(6s) \Rightarrow Ar(5p)+h\nu$ | $1.68 \times 10^7$ | calc. |
| $Ar_2^* \Rightarrow 2Ar + h\nu$ | $3.5 \times 10^5$ | [11] |

Table 6

| initial | Ar ($3s^2\ 3p^6$) | | | | | | |
|---|---|---|---|---|---|---|---|
| final | Ar(4s) | Ar(4p) | Ar(3d) | Ar(5s) | Ar(5p) | Ar(4d) | Ar(6s) |
| U (eV) | σ ($10^{-17}$) | σ ($10^{-17}$) | σ ($10^{-17}$) | σ ($10^{-18}$) | σ ($10^{-17}$) | σ ($10^{-18}$) | σ ($10^{-18}$) |
| 11 | 0 | 0 | 0 | 0 | 0 | 0 | 0 |
| 12 | 0.169 | 0 | 0 | 0 | 0 | 0 | 0 |
| 13 | 0.542 | 0.020 | 0 | 0 | 0 | 0 | 0 |
| 14 | 1.07 | 0.240 | 0 | 0 | 0 | 0 | 0 |
| 15 | 1.56 | 0.575 | 0.107 | 0.213 | 0.110 | 0.062 | 0.029 |
| 16 | 1.95 | 0.894 | 0.243 | 0.484 | 0.337 | 0.416 | 0.195 |
| 18 | 2.46 | 1.42 | 0.527 | 1.05 | 0.647 | 1.22 | 0.571 |
| 20 | 2.75 | 1.72 | 0.792 | 1.58 | 0.833 | 2.01 | 0.941 |
| 22 | 2.90 | 1.81 | 1.02 | 2.03 | 0.943 | 2.72 | 1.27 |
| 24 | 3.05 | 1.80 | 1.22 | 2.42 | 1.01 | 3.33 | 1.56 |
| 26 | 3.18 | 1.79 | 1.38 | 2.74 | 1.04 | 3.84 | 1.80 |
| 28 | 3.28 | 1.77 | 1.51 | 3.01 | 1.05 | 4.27 | 2.00 |
| 30 | 3.39 | 1.75 | 1.62 | 3.22 | 1.06 | 4.63 | 2.17 |
| 35 | 3.60 | 1.49 | 1.81 | 3.60 | 1.03 | 5.27 | 2.47 |
| 40 | 3.81 | 1.35 | 1.92 | 3.82 | 0.979 | 5.65 | 2.65 |
| 45 | 3.88 | 1.22 | 1.98 | 3.94 | 0.926 | 5.87 | 2.75 |
| 50 | 3.96 | 1.10 | 2.00 | 3.99 | 0.873 | 5.98 | 2.80 |
| 60 | 3.90 | 0.920 | 2.00 | 3.98 | 0.778 | 6.01 | 2.81 |
| 70 | 3.83 | 0.821 | 1.95 | 3.88 | 0.697 | 5.90 | 2.77 |
| 80 | 3.67 | 0.722 | 1.89 | 3.76 | 0.630 | 5.74 | 2.69 |
| 100 | 3.36 | 0.540 | 1.76 | 3.50 | 0.527 | 5.36 | 2.51 |
| 120 | 3.24 | 0.467 | 1.63 | 3.24 | 0.451 | 4.99 | 2.34 |
| 150 | 3.08 | 0.357 | 1.46 | 2.92 | 0.371 | 4.50 | 2.11 |
| 200 | 2.80 | 0.279 | 1.25 | 2.49 | 0.286 | 3.86 | 1.81 |

Table 7

| initial | Ar(4s) | Ar(4s) | Ar(4p) | Ar(4p) | Ar(4p) | Ar(4p) | Ar(3d) | Ar(5s) | Ar(5p) |
|---|---|---|---|---|---|---|---|---|---|
| final | Ar(4p) | Ar(5p) | Ar(3d) | Ar(5s) | Ar(4d) | Ar(6s) | Ar(5p) | Ar(5p) | Ar(4d,6s) |
| U(eV) | σ ($10^{-14}$) | σ ($10^{-17}$) | σ ($10^{-13}$) | σ ($10^{-14}$) | σ ($10^{-15}$) | σ ($10^{-15}$) | σ ($10^{-15}$) | σ ($10^{-13}$) | σ ($10^{-11}$) |
| 0.2 | 0 | 0 | 0 | 0 | 0 | 0 | 0 | 0 | 0 |
| 0.3 | 0 | 0 | 0 | 0 | 0 | 0 | 0 | 0 | 0.578 |
| 0.4 | 0 | 0 | 0 | 0 | 0 | 0 | 0 | 0 | 0.947 |
| 0.5 | 0 | 0 | 0 | 0 | 0 | 0 | 0 | 0 | 1.13 |
| 0.6 | 0 | 0 | 0 | 0 | 0 | 0 | 1.31 | 0.748 | 1.21 |
| 0.8 | 0 | 0 | 0 | 0 | 0 | 0 | 3.77 | 2.16 | 1.25 |
| 1.0 | 0 | 0 | 0.120 | 0.437 | 0 | 0 | 5.31 | 3.04 | 1.21 |
| 1.2 | 0 | 0 | 0.390 | 1.42 | 0 | 0 | 6.19 | 3.55 | 1.16 |
| 1.4 | 0 | 0 | 0.622 | 2.26 | 0 | 0 | 6.67 | 3.82 | 1.10 |
| 1.6 | 0.181 | 0 | 0.801 | 2.91 | 0 | 0 | 6.90 | 3.96 | 1.04 |
| 1.8 | 0.653 | 0 | 0.934 | 3.39 | 0.212 | 0.125 | 6.99 | 4.01 | 0.988 |
| 2.0 | 1.13 | 0 | 1.03 | 3.74 | 0.450 | 0.265 | 6.99 | 4.01 | 0.940 |
| 2.2 | 1.56 | 0 | 1.10 | 3.99 | 0.678 | 0.399 | 6.94 | 3.98 | 0.896 |
| 2.5 | 2.11 | 0 | 1.17 | 4.24 | 0.973 | 0.573 | 6.80 | 3.89 | 0.836 |
| 3.0 | 2.75 | 0.23 | 1.22 | 4.43 | 1.34 | 0.787 | 6.49 | 3.72 | 0.754 |
| 4.0 | 3.41 | 3.35 | 1.22 | 4.42 | 1.72 | 1.02 | 5.84 | 3.34 | 0.632 |
| 5.0 | 3.63 | 5.80 | 1.17 | 4.23 | 1.87 | 1.10 | 5.27 | 3.02 | 0.546 |
| 6.0 | 3.65 | 7.38 | 1.10 | 4.00 | 1.91 | 1.12 | 4.79 | 2.75 | 0.483 |
| 7.0 | 3.60 | 8.35 | 1.04 | 3.77 | 1.89 | 1.11 | 4.40 | 2.52 | 0.434 |
| 8.0 | 3.50 | 8.92 | 0.98 | 3.56 | 1.85 | 1.09 | 4.07 | 2.33 | 0.394 |
| 9.0 | 3.39 | 9.23 | 0.926 | 3.36 | 1.80 | 1.06 | 3.79 | 2.17 | 0.362 |
| 10 | 3.28 | 9.39 | 0.878 | 3.19 | 1.74 | 1.03 | 3.54 | 2.03 | 0.335 |
| 11 | 3.17 | 9.43 | 0.835 | 3.03 | 1.69 | 0.993 | 3.33 | 1.91 | 0.312 |
| 12 | 3.06 | 9.40 | 0.795 | 2.89 | 1.63 | 0.960 | 3.15 | 1.80 | 0.293 |
| 13 | 2.95 | 9.33 | 0.760 | 2.76 | 1.58 | 0.929 | 2.98 | 1.71 | 0.275 |
| 14 | 2.85 | 9.22 | 0.727 | 2.64 | 1.53 | 0.899 | 2.84 | 1.63 | 0.260 |
| 15 | 2.76 | 9.09 | 0.698 | 2.53 | 1.48 | 0.871 | 2.71 | 1.55 | 0.247 |
| 16 | 2.67 | 8.95 | 0.671 | 2.43 | 1.43 | 0.844 | 2.59 | 1.48 | 0.235 |
| 18 | 2.51 | 8.65 | 0.623 | 2.26 | 1.35 | 0.795 | 2.38 | 1.37 | 0.215 |
| 20 | 2.37 | 8.35 | 0.582 | 2.11 | 1.28 | 0.751 | 2.21 | 1.27 | 0.198 |
| 22 | 2.25 | 8.05 | 0.547 | 1.98 | 1.21 | 0.712 | 2.06 | 1.18 | 0.183 |
| 24 | 2.13 | 7.76 | 0.516 | 1.87 | 1.15 | 0.677 | 1.94 | 1.11 | 0.171 |
| 26 | 2.03 | 7.49 | 0.488 | 1.77 | 1.10 | 0.646 | 1.82 | 1.05 | 0.161 |
| 28 | 1.94 | 7.23 | 0.464 | 1.68 | 1.05 | 0.617 | 1.73 | 0.990 | 0.151 |
| 30 | 1.86 | 6.99 | 0.442 | 1.61 | 1.00 | 0.591 | 1.64 | 0.940 | 0.143 |
| 35 | 1.68 | 6.45 | 0.397 | 1.44 | 0.910 | 0.536 | 1.46 | 0.836 | 0.127 |
| 40 | 1.54 | 5.99 | 0.360 | 1.31 | 0.834 | 0.491 | 1.32 | 0.755 | 0.114 |
| 45 | 1.42 | 5.59 | 0.330 | 1.20 | 0.770 | 0.453 | 1.20 | 0.689 | 0.103 |
| 50 | 1.32 | 5.25 | 0.305 | 1.11 | 0.716 | 0.422 | 1.11 | 0.635 | 0.094 |
| 60 | 1.16 | 4.68 | 0.266 | 0.966 | 0.630 | 0.371 | 0.960 | 0.550 | 0.081 |
| 70 | 1.04 | 4.23 | 0.237 | 0.859 | 0.564 | 0.332 | 0.849 | 0.486 | 0.071 |
| 80 | 0.94 | 3.87 | 0.214 | 0.775 | 0.511 | 0.301 | 0.763 | 0.437 | 0.064 |
| 100 | 0.795 | 3.32 | 0.179 | 0.651 | 0.433 | 0.255 | 0.637 | 0.365 | 0.053 |
| 120 | 0.692 | 2.91 | 0.155 | 0.563 | 0.377 | 0.222 | 0.549 | 0.314 | 0.045 |
| 150 | 0.582 | 2.47 | 0.130 | 0.471 | 0.317 | 0.187 | 0.456 | 0.262 | 0.038 |
| 200 | 0.464 | 1.99 | 0.103 | 0.373 | 0.253 | 0.149 | 0.359 | 0.206 | 0.029 |

Table 8

| initial | Ar | Ar(4s) | Ar(4p) | Ar(3d) | Ar(5s) | Ar(5p) | Ar(4d) | Ar(6s) |
|---|---|---|---|---|---|---|---|---|
| final | $Ar^+$ | | | | | | | |
| U (eV) | σ (10⁻¹⁶) | σ (10⁻¹⁵) | σ (10⁻¹⁴) | σ (10⁻¹⁴) | σ (10⁻¹⁴) | σ (10⁻¹⁴) | σ (10⁻¹³) | σ (10⁻¹⁴) |
| 0.8 | 0 | 0 | 0 | 0 | 0 | 0 | 0 | 0 |
| 1.0 | 0 | 0 | 0 | 0 | 0 | 0 | 0.182 | 0.608 |
| 1.2 | 0 | 0 | 0 | 0 | 0 | 0.87 | 0.829 | 2.76 |
| 1.4 | 0 | 0 | 0 | 0 | 0 | 4.13 | 1.20 | 4.01 |
| 1.6 | 0 | 0 | 0 | 0 | 0 | 6.18 | 1.43 | 4.75 |
| 1.8 | 0 | 0 | 0 | 1.16 | 0.387 | 7.51 | 1.56 | 5.20 |
| 2.0 | 0 | 0 | 0 | 2.40 | 0.801 | 8.39 | 1.64 | 5.47 |
| 2.2 | 0 | 0 | 0 | 3.30 | 1.10 | 8.97 | 1.69 | 5.63 |
| 2.5 | 0 | 0 | 0 | 4.22 | 1.41 | 9.48 | 1.72 | 5.72 |
| 3.0 | 0 | 0 | 0.823 | 5.10 | 1.70 | 9.79 | 1.70 | 5.67 |
| 4.0 | 0 | 0 | 1.64 | 5.70 | 1.90 | 9.54 | 1.59 | 5.29 |
| 5.0 | 0 | 1.44 | 1.95 | 5.71 | 1.90 | 8.95 | 1.45 | 4.84 |
| 6.0 | 0 | 2.27 | 2.06 | 5.53 | 1.84 | 8.32 | 1.33 | 4.43 |
| 7.0 | 0 | 2.72 | 2.08 | 5.29 | 1.76 | 7.73 | 1.22 | 4.07 |
| 8.0 | 0 | 2.97 | 2.05 | 5.03 | 1.68 | 7.19 | 1.12 | 3.75 |
| 9.0 | 0 | 3.10 | 2.01 | 4.77 | 1.59 | 6.71 | 1.04 | 3.47 |
| 10 | 0 | 3.17 | 1.95 | 4.53 | 1.51 | 6.28 | 0.970 | 3.23 |
| 11 | 0 | 3.18 | 1.89 | 4.31 | 1.44 | 5.90 | 0.907 | 3.02 |
| 12 | 0 | 3.17 | 1.83 | 4.10 | 1.37 | 5.57 | 0.852 | 2.84 |
| 13 | 0 | 3.14 | 1.77 | 3.91 | 1.30 | 5.26 | 0.802 | 2.67 |
| 14 | 0 | 3.10 | 1.71 | 3.73 | 1.24 | 4.99 | 0.758 | 2.53 |
| 15 | 0 | 3.05 | 1.65 | 3.57 | 1.19 | 4.74 | 0.719 | 2.40 |
| 16 | 0.020 | 3.00 | 1.59 | 3.42 | 1.14 | 4.52 | 0.683 | 2.28 |
| 18 | 0.294 | 2.88 | 1.49 | 3.16 | 1.05 | 4.12 | 0.621 | 2.07 |
| 20 | 0.627 | 2.76 | 1.40 | 2.93 | 0.975 | 3.79 | 0.569 | 1.90 |
| 22 | 0.933 | 2.65 | 1.32 | 2.73 | 0.909 | 3.51 | 0.525 | 1.75 |
| 24 | 1.18 | 2.54 | 1.25 | 2.55 | 0.851 | 3.27 | 0.488 | 1.63 |
| 26 | 1.41 | 2.44 | 1.18 | 2.40 | 0.800 | 3.06 | 0.455 | 1.52 |
| 28 | 1.60 | 2.34 | 1.12 | 2.26 | 0.754 | 2.87 | 0.427 | 1.42 |
| 30 | 1.80 | 2.25 | 1.07 | 2.14 | 0.713 | 2.70 | 0.402 | 1.34 |
| 35 | 2.17 | 2.05 | 0.950 | 1.88 | 0.628 | 2.36 | 0.350 | 1.17 |
| 40 | 2.39 | 1.87 | 0.858 | 1.68 | 0.561 | 2.10 | 0.310 | 1.03 |
| 45 | 2.49 | 1.73 | 0.781 | 1.52 | 0.507 | 1.89 | 0.278 | 0.927 |
| 50 | 2.53 | 1.60 | 0.717 | 1.39 | 0.462 | 1.72 | 0.252 | 0.841 |
| 60 | 2.66 | 1.39 | 0.615 | 1.18 | 0.393 | 1.45 | 0.213 | 0.709 |
| 70 | 2.77 | 1.23 | 0.539 | 1.03 | 0.342 | 1.26 | 0.184 | 0.613 |
| 80 | 2.84 | 1.11 | 0.479 | 0.907 | 0.302 | 1.11 | 0.162 | 0.540 |
| 100 | 2.85 | 0.918 | 0.392 | 0.737 | 0.246 | 0.895 | 0.131 | 0.436 |
| 120 | 2.81 | 0.784 | 0.332 | 0.620 | 0.207 | 0.751 | 0.110 | 0.365 |
| 150 | 2.68 | 0.643 | 0.269 | 0.501 | 0.167 | 0.605 | 0.882 | 0.294 |
| 200 | 2.50 | 0.494 | 0.205 | 0.380 | 0.127 | 0.457 | 0.666 | 0.222 |

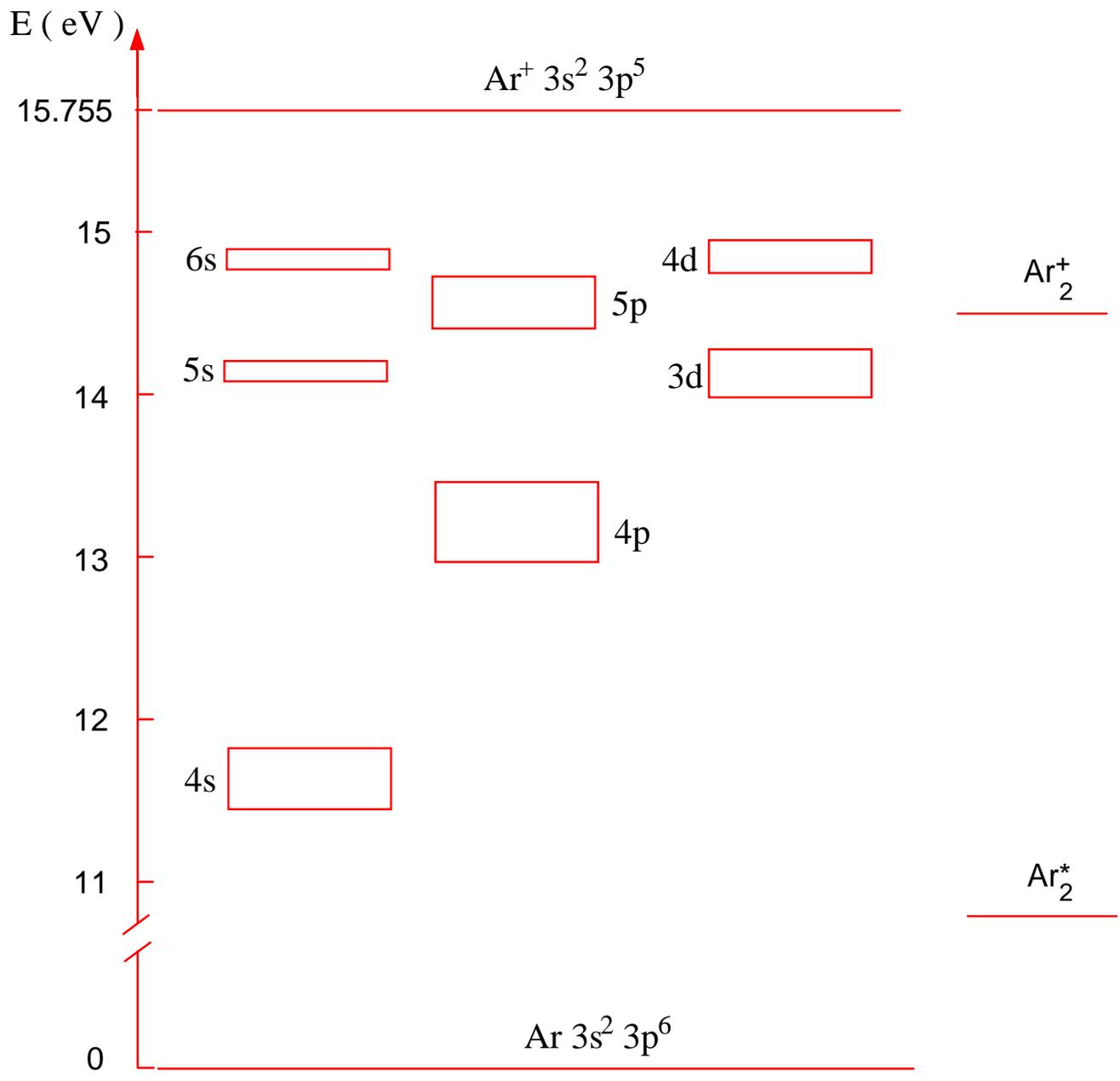

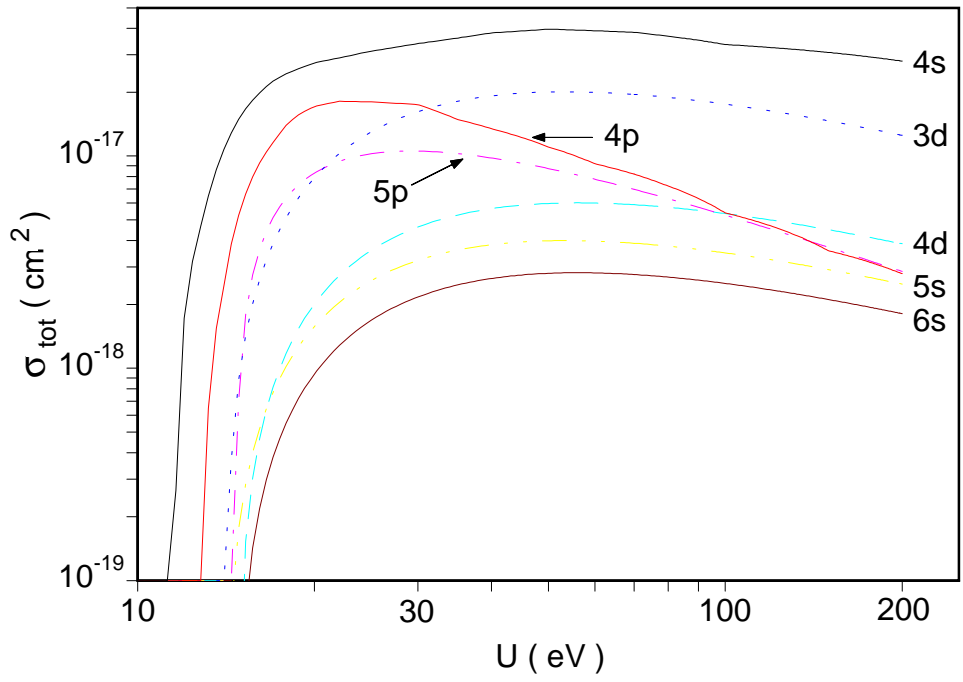

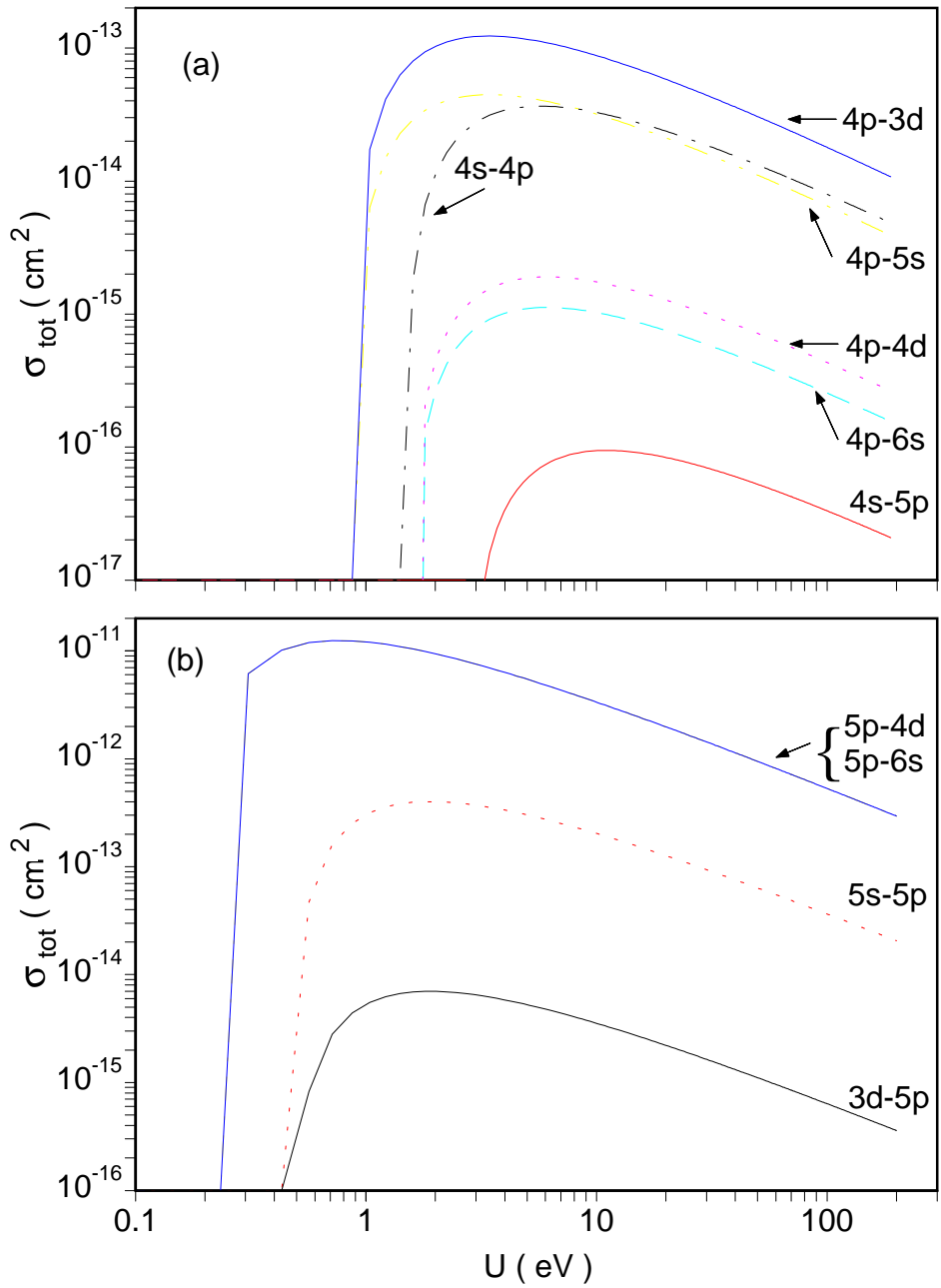

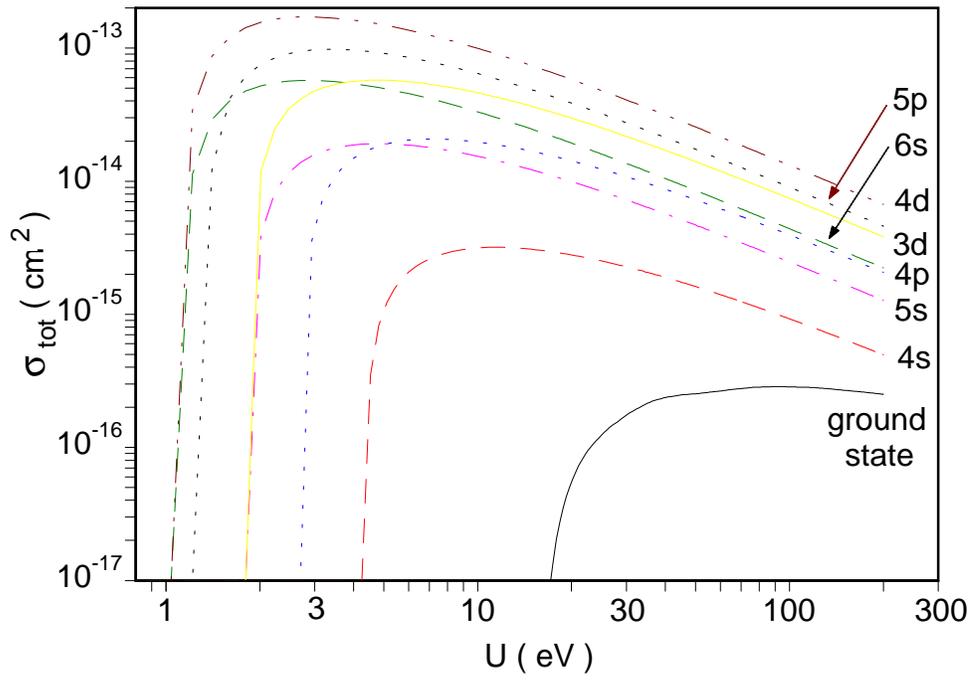

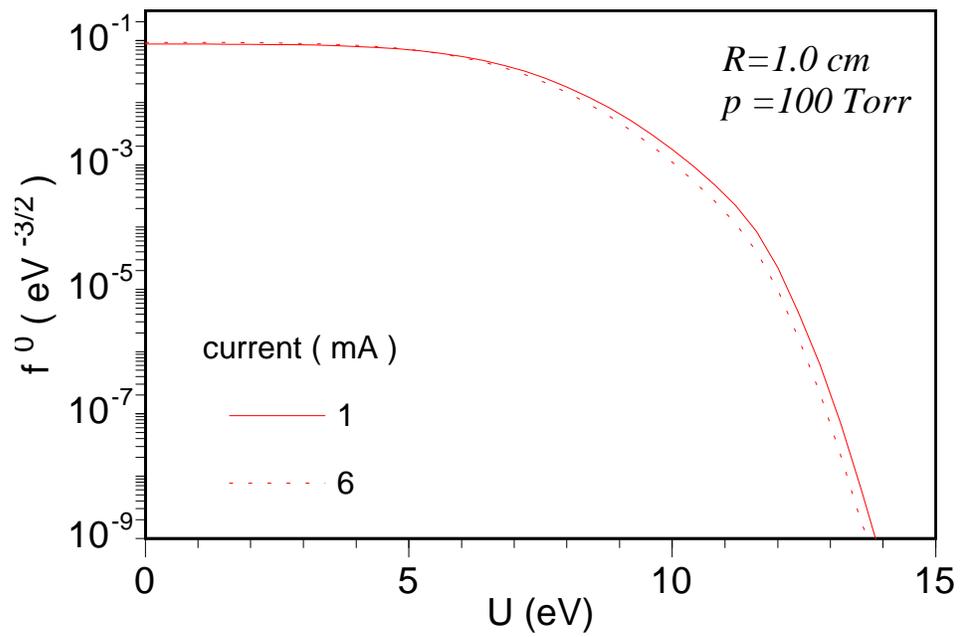

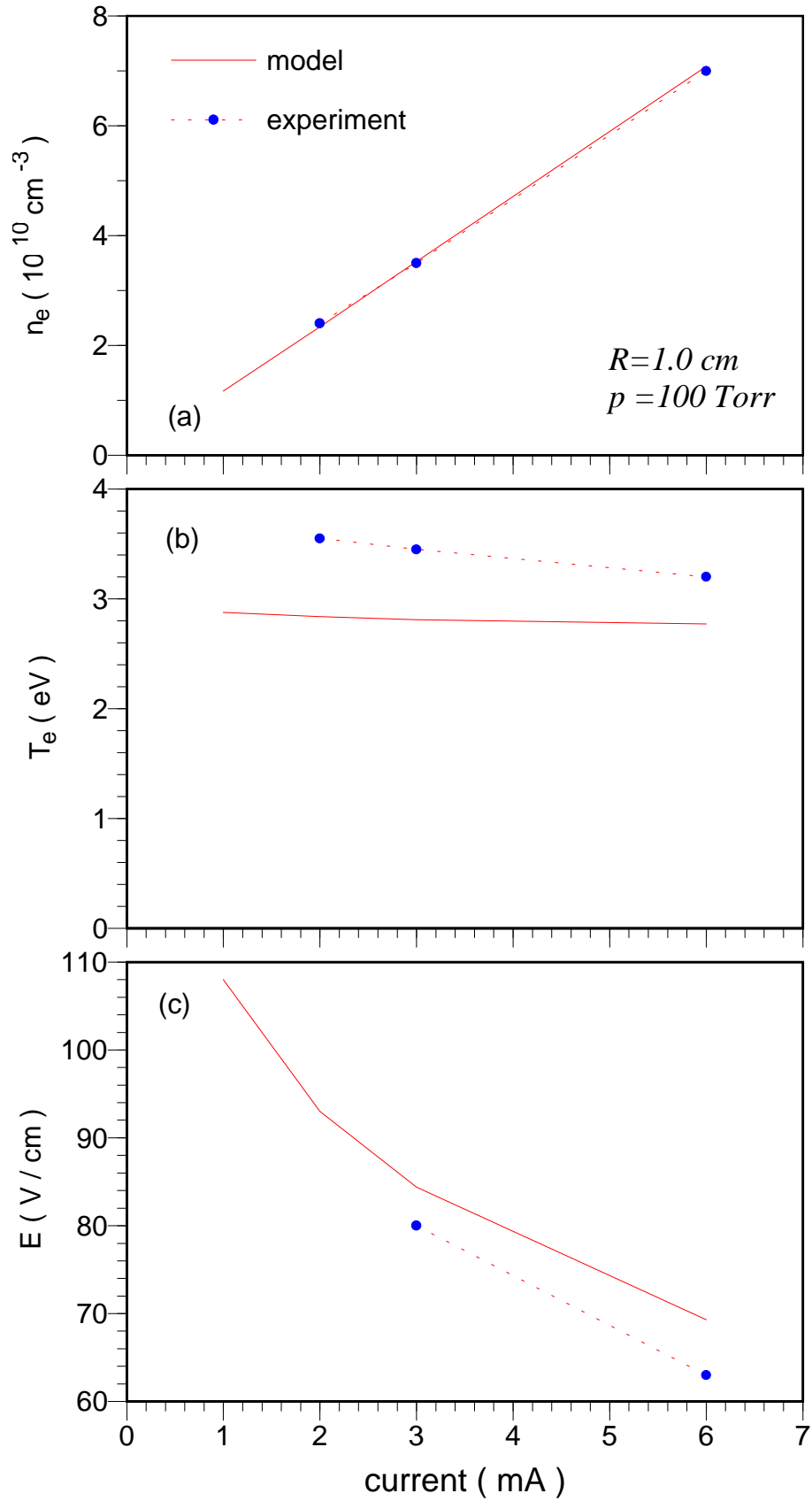

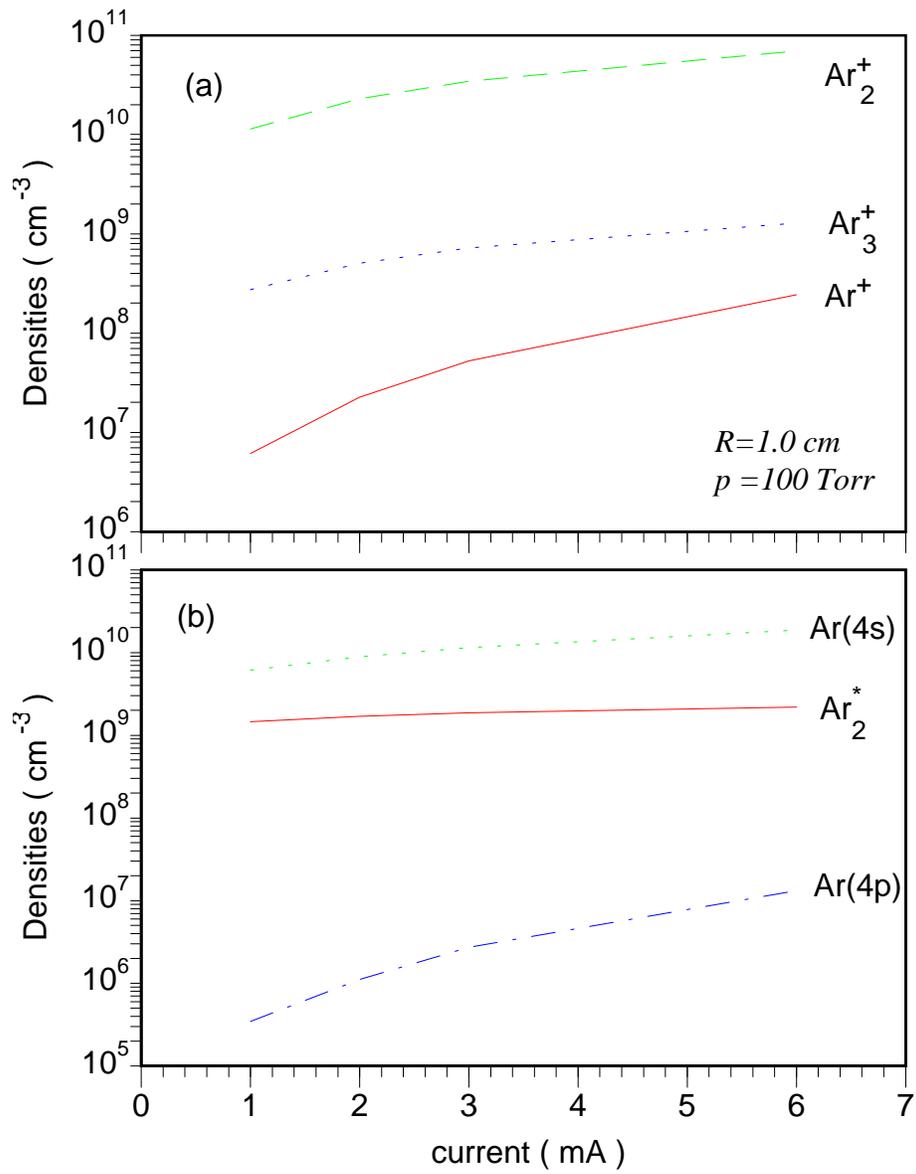

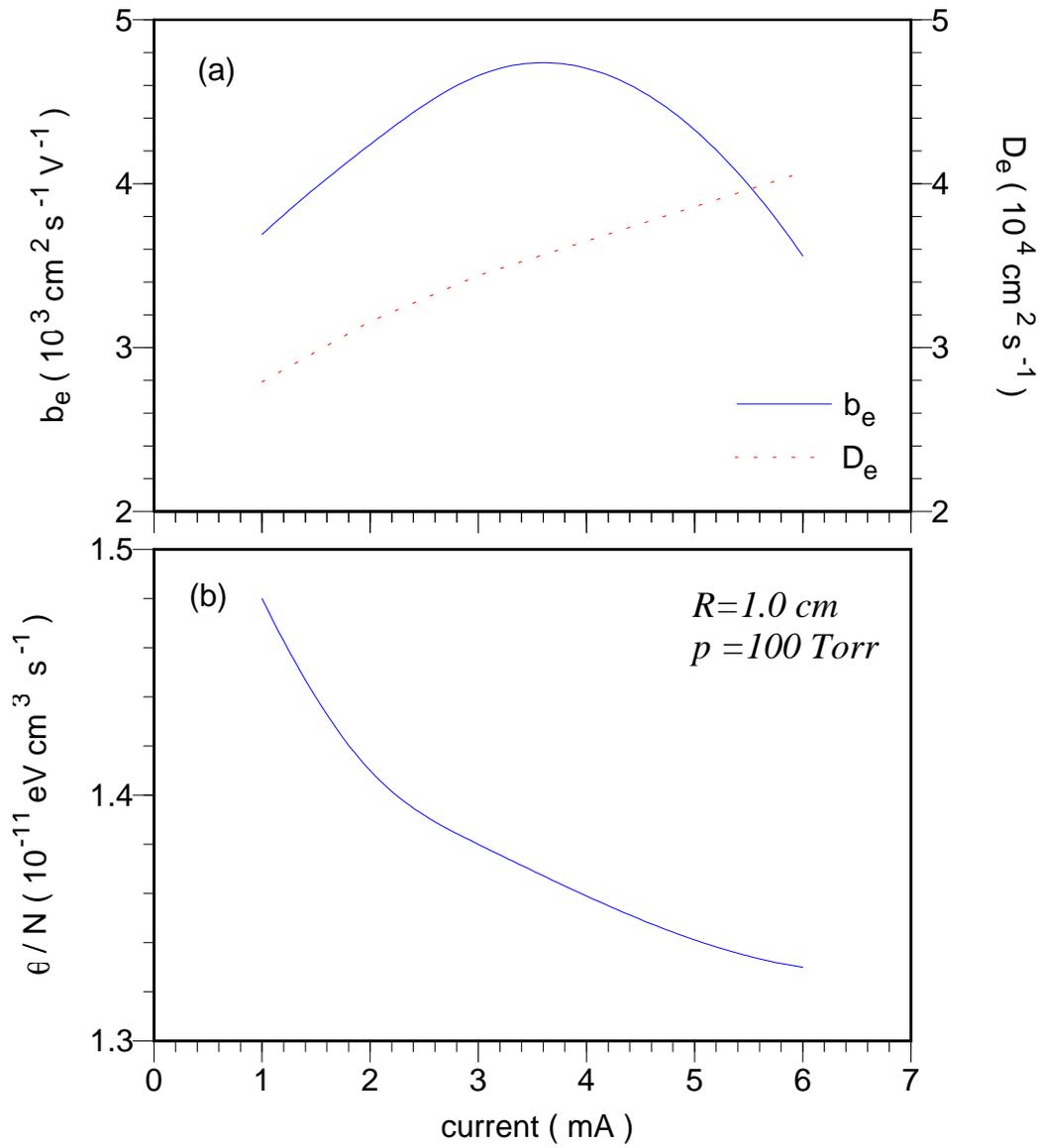

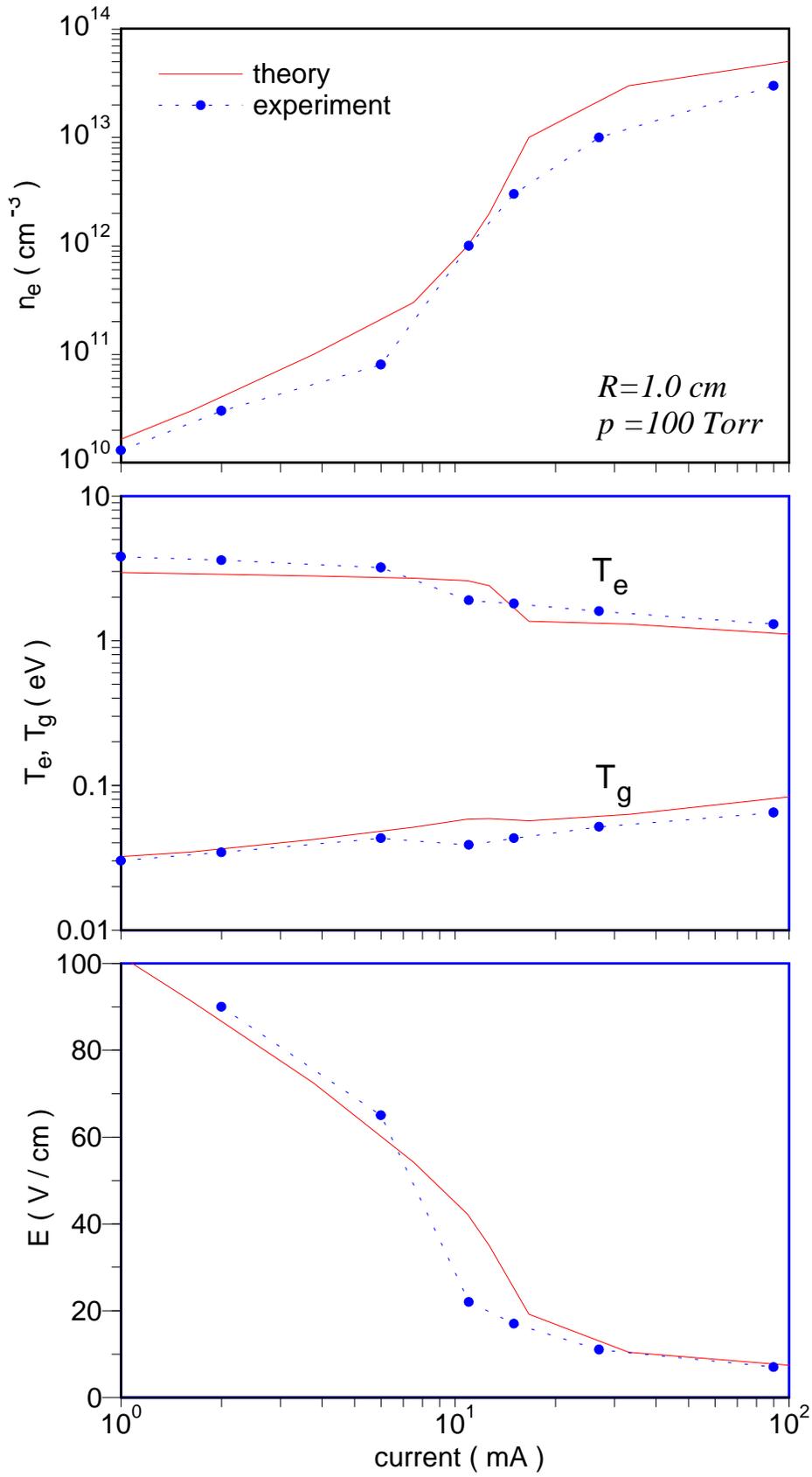

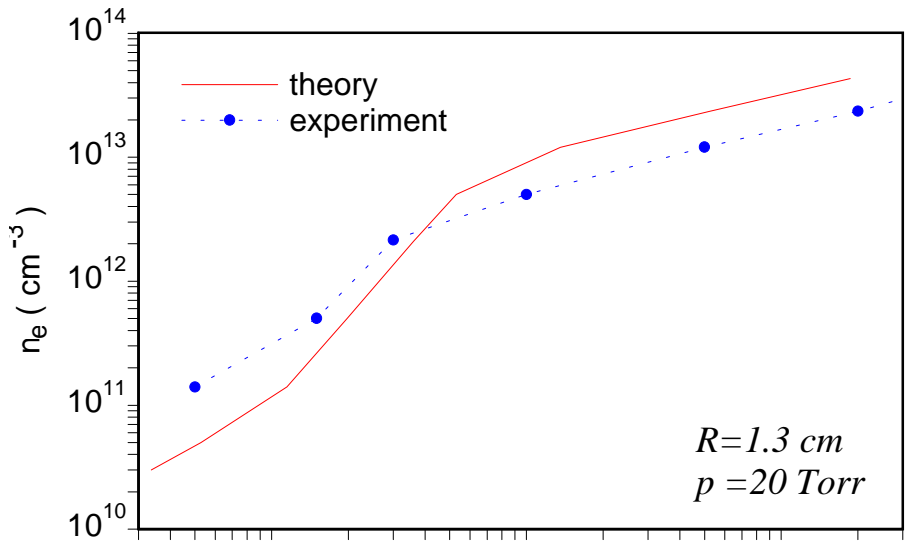
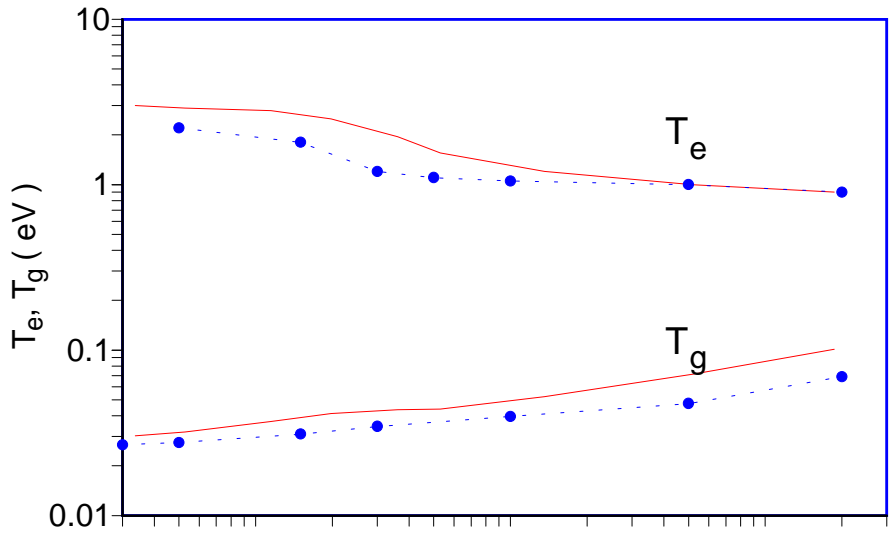
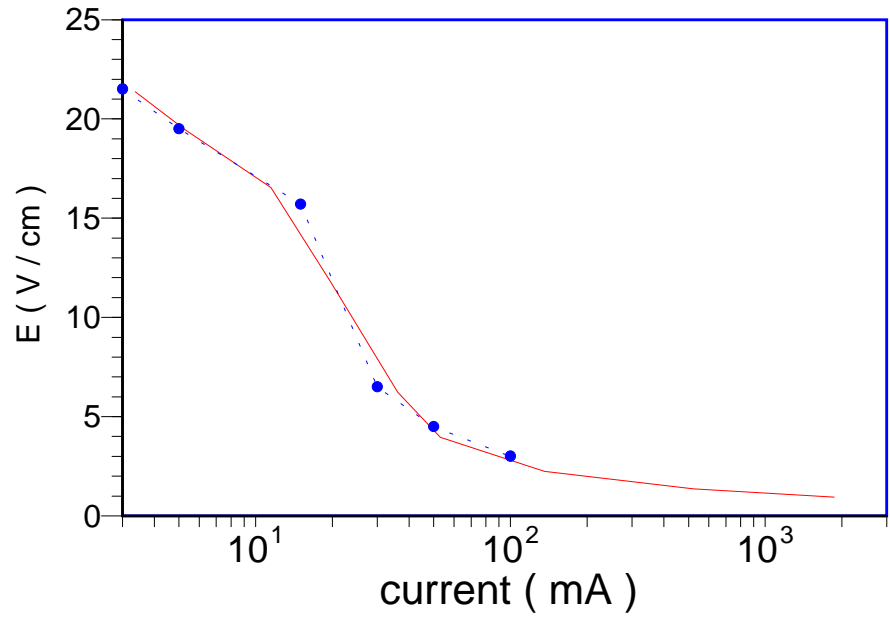